\documentclass[
    final            % use final for the camera ready runs
  ]
  {aipproc}

\layoutstyle{6x9}
\usepackage[absolute]{textpos}

\begin{document}

\title{Development of cluster-jet targets: From COSY-11 to FAIR.}

%http://www.aip..org/pacs/index.html

\author{A. Täschner}{
  address={Institut für Kernphysik, Wilhelm-Klemm-Str. 9, D-48149 Münster, Germany}
}

\author{S. General}{
  address={Institut für Kernphysik, Wilhelm-Klemm-Str. 9, D-48149 Münster, Germany}
}

%\author{H.-W. Ortjohann}{
%  address={Institut für Kernphysik, Wilhelm-Klemm-Str. 9, D-48149 Münster, Germany}
%}

\author{J. Otte}{
  address={Institut für Kernphysik, Wilhelm-Klemm-Str. 9, D-48149 Münster, Germany}
}

\author{T. Rausmann}{
  address={Institut für Kernphysik, Wilhelm-Klemm-Str. 9, D-48149 Münster, Germany}
}

\author{A. Khoukaz}{
  address={Institut für Kernphysik, Wilhelm-Klemm-Str. 9, D-48149 Münster, Germany}
}

\begin{abstract}
The development of cluster-jet targets of Münster type is presented. Starting
with the first target installed at the COSY-11 experiment the progress is
described which was made at a cluster-jet target facility installed in Münster leading
to a prototype for a cluster-jet target for the upcoming PANDA experiment at FAIR.
\end{abstract}

\maketitle

%%%%%%%%%%%%%%%%%%%%%%%%%%%%%%%%%%%%%%%%%%%%
%% MAINMATTER
%%%%%%%%%%%%%%%%%%%%%%%%%%%%%%%%%%%%%%%%%%%%

\section{The cluster-jet target for COSY-11}

The COSY-11 experiment \cite{Brauksiepe1996} is an internal storage ring experiment
at the cooler synchrotron COSY at the FZ Jülich.
Its main purpose was the investigation of the meson production close
to the production threshold in proton-proton and proton-deuteron collisions.

Several design goals had to be achieved during the development of  the target for COSY-11. At first
the target had to provide a small and precisely known interaction region with
the accelerator beam in order to allow for a precise momentum reconstruction
of the ejectiles. This implied directly that the influence of the vacuum
in the storage ring had to be small and confined to the target region.
The target thickness had to be small enough in order to achieve beam lifetimes in
the order of 10 min to 1h but on the other hand it had to be large enough in order to provide
acceptably high event rates. These constraints ruled out the use of gas-jet targets or solid
targets leading to the decision to build a cluster-jet target. In such targets
hydrogen or deuterium gas is cooled down to temperatures of \textasciitilde30 K and is
pressed through a thin laval nozzle. The gas expands into a vaccuum chamber where a
small fraction of the gas condensates to nanoparticles called clusters.
Using several skimmers a cluster-jet is formed which traverses
the accelerator beam with a small angular divergence given by the skimmer
geometry. 

Since the target facility had to be installed into a gap of 30 cm between a dipole
and a quadrupole magnet of the storage ring, it was not possible to use
cluster-jet designs which were in operation at other storage rings at that time.
The cluster-jet target of the Münster type \cite{Dombrowski1997a}
uses an innovative pumping scheme with specially designed cryopumps in order
to meet these space requirements. After the construction and successful tests
in Münster it was transferred to Jülich where it was reassembled. In 1995
it was installed at the COSY-11 place where it worked very reliable 
and met all demands up to the end of the experiment in 2006. 
\newlength{\oldparindent}
\setlength{\oldparindent}{\parindent}
\setlength{\parindent}{0pt}
\begin{textblock*}{140mm}(38mm,10mm)
\footnotesize
Copyright (2007) American Institute of Physics. This article may be downloaded
for personal use only. Any other use requires prior permission of the author
and the American Institute of Physics. The following article appeared in
AIP Conf. Proc. \textbf{950}, 85 (2007) and may be found at
http://dx.doi.org/10.1063/1.2819057.
\end{textblock*}
\setlength{\parindent}{\oldparindent}
\section{Cluster-jet target facilty in Münster}

\begin{table}
\begin{tabular}{lccccc}
\hline
  & \tablehead{1}{c}{b}{CELSIUS}
  & \tablehead{1}{c}{b}{E835\\FERMILAB}
  & \tablehead{1}{c}{b}{ANKE\\COSY-11}
  & \vspace{2mm} & \tablehead{1}{c}{b}{Münster}\\
\hline
nozzle diameter & 100 \textmu m & 37 \textmu m & 11-16 \textmu m &  & 16-28 \textmu m \\
gas temperature & 20-35 K & 20-32 K & 22-35 K & & 20-35 K \\
gas pressure & 1.4 bar & < 10 bar & 18 bar & & > 18 bar \\
distance from nozzle & 0.32 m & 0.26 m & 0.65 m & & \textbf{2.1 m} \\
target thickness & $1.3\cdot10^{14}\ \mathrm{cm}^{-2}$ & 
$3\cdot10^{14}\ \mathrm{cm}^{-2}$ & > $10^{14}\ \mathrm{cm}^{-2}$ & 
& > $4\cdot10^{14}\ \mathrm{cm}^{-2}$\\
\hline
\end{tabular}
\caption{Typical operation parameters of hydrogen cluster-jet targets.}
\label{tab:parameters}
\end{table}

\begin{figure}
  \includegraphics[width=.63\textwidth]{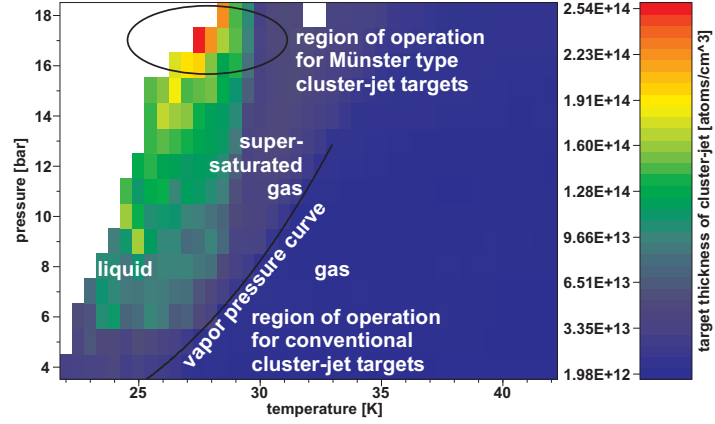}
  \caption{Target thickness of the target facility in Münster as
  function of the gas pressure and temperature. Shown are also the 
  regions of operation for conventional targets compared to the Münster type.}
  \label{fig:mapping}
\end{figure}

Since the distance~$d$ between between nozzle and interaction point 
is connected to the target thickness~$n_T$ via the formula $n_T\propto d^{-2}$ it is 
remarkable that the target thickness of targets listed in the first three columns of table 
\ref{tab:parameters} are in the same range although the distance at  ANKE and COSY-11 
is enlarged by a factor of two.
This shows the advantages of the Münster type targets which are operated in a regime 
where the hydrogen is supersaturated before the nozzle (see figure \ref{fig:mapping}) 
leading to a drastic increase in target thickness.

\begin{figure}
  \includegraphics[width=.6\textwidth]{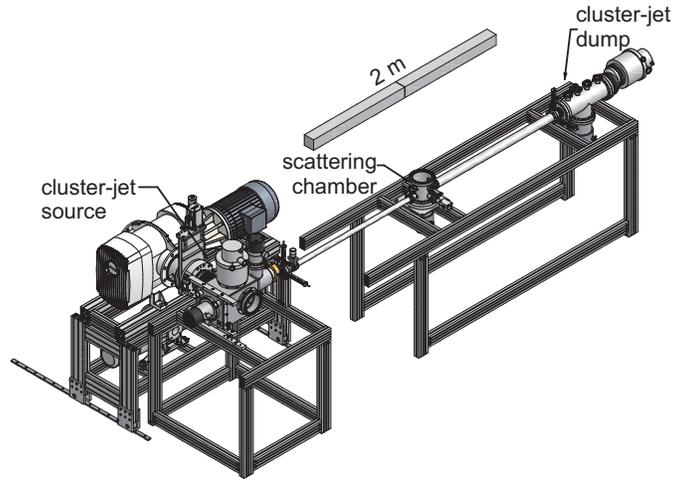}
  \caption{CAD drawing of the cluster-jet target used in Münster.}
  \label{fig:MuensterCAD}
\end{figure}

Future experiments in hadron physics like the $\bar{\mathrm{P}}$ANDA experiment at FAIR \cite{PANDATPR}
will use 4$\pi$-detectors leading to a drastic increase in the distance between target
and interaction point. In the case of the $\bar{\mathrm{P}}$ANDA 
experiment the distance will be approximately two meters. In order to 
have the same target thickness as used in the established
experiments it is necessary to increase the target thickness by at least one order
of magnitude.
In order to test the feasibility of such an increase in target thickness 
a complete target facility was build up in Münster which has a distance of 2.1 m between 
nozzle and interaction point. This distance represents the worst case scenario for
the $\bar{\mathrm{P}}$ANDA detector. Figure~\ref{fig:MuensterCAD} shows a CAD drawing
of the current setup in Münster. The cluster-jet produced in the cluster source
traverses a vacuum chamber which is placed at the distance where the beam-target interaction
would take place in the PANDA detector and is finally collected and pumped in the beam dump.

There are two parameters which can be changed when the cluster-jet target
is in operation, namely the gas temperature and gas pressure before the nozzle.
In figure~\ref{fig:mapping} the target thickness measured in the scattering
chamber is shown as function of these two parameters. 
The target thickness increases with increasing pressure and decreasing temperature.
After detailed studies several improvements had been made in order to be able to operate 
at the point of highest thickness, e.g. the pumping speeds at the collimator and 
the skimmer chamber have been increased and the gas system has been modified to reduce heat losses.

The target thickness is measured by inserting a small rod into the cluster-jet beam
stopping the clusters colliding with the rod which leads to a pressure increase
in the scattering chamber. From the measured pressure increase the cluster-jet flow~$\Phi$
can be calculated. Together with the cross section~$A$ between rod and cluster-jet and
the velocity~$v_{jet}$ the areal target thickness~$n_T$ can be calculated

\begin{equation}
	n_T=\frac{\Phi}{A\cdot v_{jet}}.
	\label{eq:targetthickness}
\end{equation}

\begin{figure}
  \includegraphics[width=.62\textwidth]{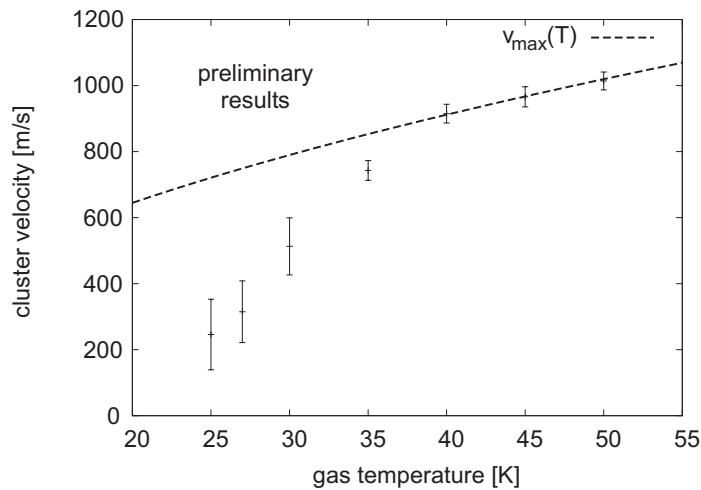}
  \caption{Velocity of the clusters measured using a TOF technique as function of the gas temperature 
  at a constant gas pressure of 8 bar.}
  \label{fig:vCluster}
\end{figure}

Up to the end of last year the cluster velocity was calculated from the theoretically estimated
maximum velocity of the gas leaving the laval nozzle. 
Measurements performed at the E835 experiment at FERMILAB confirmed this
assumption \cite{Allspach1998} for the operation regime in the gaseous part
above the vapor pressure curve. Since the Münster type target operates
at much higher pressures and lower temperatures it had to be tested whether
this assumptions holds also for the target setup used in Münster.
Therefore, velocity measurements were performed using a time-of-flight technique
where the clusters were ionized using a pulsed electron gun situated
near the exit of the cluster source and detected with a channeltron at
the end of the cluster-jet dump. First preliminary results of these measurements
are show in figure \ref{fig:vCluster} where the cluster velocity is plotted
as function of the gas temperature at a constant gas pressure. 
In the region above the vapor pressure curve the measured velocity 
agrees well with the maximum gas velocity, but at lower
temperatures the measured velocity is by a factor 2 to 3 lower than the prediction.
This leads to an increase of the real target thickness by the same factor.

The maximum target thickness shown in figure~\ref{fig:mapping}
was calculated using the maximum gas velocity as cluster velocity.
Considering the measured velocities the real target density at the interaction point
turns out to be larger by a factor of 2-3. Due to this a target thickness
of at least $4\cdot10^{14}$ atoms/cm$^2$ was achieved at a distance of 2.1 m 
between nozzle and interaction point. Further density increases are currently limited
by the finite cooling power of the present setup.

\section{Prototype for $\bar{\mathrm{P}}$ANDA}

After achieving the first goal of increasing the cluster-jet target thickness by a factor of 10
a new cluster source is currently build up which should be able to achieve lower temperatures
and therefore higher target thicknesses. This new setup is at the same time
a prototype for a cluster-jet target which can be installed at $\bar{\mathrm{P}}$ANDA.
The new setup will have a much more powerful coldhead for the gas cooling and
will have improvements in the heat insulations. It will be possible to move the
skimmer and the collimator in order to optimize the beam optics. Densities of $10^{15}$ atoms/cm$^2$
in combination with $\bar{\mathrm{P}}$ANDA geometry are expected

%%%%%%%%%%%%%%%%%%%%%%%%%%%%%%%%%%%%%%%%%%%%%%%%
%% BACKMATTER
%%%%%%%%%%%%%%%%%%%%%%%%%%%%%%%%%%%%%%%%%%%%%%%%

%\begin{theacknowledgments}
%We acknowledge the support of the EU FP6, of the BMBF, and  of the FZJ FFE. 
%\end{theacknowledgments}



\bibliographystyle{aipproc}   % if natbib is available

\bibliography{DCJTK07}

\end{document}